\documentclass[prb,superscriptaddress,twocolumn,floatfix,amsmath,amssymb,showpacs]{revtex4}

\usepackage{graphicx}
\usepackage{dcolumn}
\usepackage{bm}

\begin{document}

\title{Novel weak-ferromagnetic metallic state in heavily doped Ba$_{1-x}$K$_{x}$Mn$_{2}$As$_{2}$}

\author{Jin-Ke Bao}
\affiliation{Department of Physics, Zhejiang University, Hangzhou
310027, China}

\author{Hao Jiang}
\affiliation{Department of Physics, Zhejiang University, Hangzhou
310027, China}

\author{Yun-Lei Sun}
\affiliation{Department of Physics, Zhejiang University, Hangzhou
310027, China}

\author{Wen-He Jiao}
\affiliation{Department of Physics, Zhejiang University, Hangzhou
310027, China}

\author{Chen-Yi Shen}
\affiliation{Department of Physics, Zhejiang University, Hangzhou
310027, China}

\author{Han-Jie Guo}
\affiliation{Department of Physics, Zhejiang University, Hangzhou
310027, China}

\author{Ye Chen}
\affiliation{Department of Physics, Zhejiang University, Hangzhou
310027, China}

\author{Chun-Mu Feng}
\affiliation{Department of Physics, Zhejiang University, Hangzhou
310027, China}

\author{Hui-Qiu Yuan}
\affiliation{Department of Physics, Zhejiang University, Hangzhou
310027, China}

\author{Zhu-An Xu}
\affiliation{Department of Physics, Zhejiang University, Hangzhou
310027, China}

\author{Guang-Han Cao}
\email[corresponding author: ]{ghcao@zju.edu.cn}
\affiliation{Department of Physics, Zhejiang University, Hangzhou
310027, China} \affiliation{Institute for Solid State Physics, The
University of Tokyo, Kashiwanoha, Kashiwa, Chiba 277-8581, Japan}

\author{Ryo Sasaki}
\affiliation{Institute for Solid State Physics, The University of
Tokyo, Kashiwanoha, Kashiwa, Chiba 277-8581, Japan}

\author{Toshiki Tanaka}
\affiliation{Institute for Solid State Physics, The University of
Tokyo, Kashiwanoha, Kashiwa, Chiba 277-8581, Japan}

\author{Kazuyuki Matsubayashi}
\affiliation{Institute for Solid State Physics, The University of
Tokyo, Kashiwanoha, Kashiwa, Chiba 277-8581, Japan}

\author{Yoshiya Uwatoko}
\affiliation{Institute for Solid State Physics, The University of
Tokyo, Kashiwanoha, Kashiwa, Chiba 277-8581, Japan}

\begin{abstract}
Heavily doped Ba$_{1-x}$K$_{x}$Mn$_{2}$As$_{2}$ ($x$=0.19 and 0.26)
single crystals were successfully grown, and investigated by the
measurements of resistivity and anisotropic magnetization. In
contrast to the antiferromagnetic insulating ground state of the
undoped BaMn$_{2}$As$_{2}$, the K-doped crystals show metallic
conduction with weak ferromagnetism below $\sim$50 K and
Curie-Weiss-like in-plane magnetic susceptibility above $\sim$50 K.
Under high pressures up to 6 GPa, the low-temperature metallicity
changes into a state characterized by a Kondo-like resistivity
minimum. Electronic structure calculations for $x$=0.25 using
$2\times2\times1$ supercell reproduce the hole-doped metallic state.
The density of states at Fermi energy have significant As 4$p$
components, suggesting that the 4$p$ holes are mainly responsible
for the metallic conduction. Our results suggest that the interplay
between itinerant 4$p$ holes and local 3$d$ moments are mostly
responsible for the novel metallic state.
\end{abstract}



\maketitle
\section{\label{sec:level1}Introduction}
The discovery of Fe-based pnictide
superconductors\cite{hosono,review} has also brought considerable
research interest on the isostructural pnictides containing other
transition metals, like La$T_M$AsO and Ba$T_M$$_{2}$As$_{2}$ ($T_M$
stands for 3$d$ transition metals such as Mn, Fe, Co and Ni). Among
them, the Mn-based materials are particularly worthy of
investigation, because they exhibit antiferromagnetic (AFM)
\emph{insulating} ground state, similar to the parent compounds of
high-$T_c$ cuprates and colossal magnetoresistance manganites.
According to the paradigm of metal-insulator transition,\cite{mit}
it is expected to realize an insulator-to-metal transition either by
charge carrier doping (band filling control) or by applying high
pressures (band-width control) in the Mn-based pnictides. Very
recently Tokura and co-workers\cite{tokura} indeed observed this
kind of transition via electron doping in oxygen-deficient
SmMnAsO$_{1-\delta}$. Coincidentally we realized the metallization
by hole doping in La$_{1-x}$Sr$_{x}$MnAsO.\cite{syl} Notably the
metallic La$_{0.9}$Sr$_{0.1}$MnAsO polycrystalline sample showed an
unusually large Seebeck coefficient of $\sim240 \mu$V/K.

Although the Ba$T_M$$_{2}$As$_{2}$ family all crystallize in
elongated ThCr$_{2}$Si$_{2}$-type structure, they show diverse
physical properties. BaFe$_{2}$As$_{2}$ is a striped AFM semimetal
with an ordered moment of $\sim $0.9$ \mu_{B}$/Fe,\cite{BaFe2As2}
serving as a prototype parent compound of Fe-based
superconductors.\cite{122Fe} Ba$_{2}$Cr$_{2}$As$_{2}$ was found to
be a $G$-type AFM metal with an ordered moment of $\sim $2$
\mu_{B}$/Cr and with strong Cr 3$d-$As 4$p$
hybridization.\cite{122Cr} BaCo$_{2}$As$_{2}$ exhibits a
renormalized paramagnetic metal in proximity to a possible
ferromagnetic quantum critical point.\cite{122Co} BaNi$_{2}$As$_{2}$
was demonstrated as a fully-gapped superconductor.\cite{122Ni} It
seems that the count of 3$d$ electrons determines the physical
properties, and the metallic states can be understood by the partial
filling of 3$d$ electrons. Exceptionally BaMn$_{2}$As$_{2}$ has a
$G$-type AFM \emph{insulating} ground state with a large moment
($\sim$3.9 $\mu_B$/Mn) and a high N\'{e}el temperature ($\sim$620
K).\cite{122Mn,johnston-prop,johnston-nd,band} The magnetic
properties can be well described by the $J_{1}-J_{2}-J_{c}$
Heisenberg exchange model.\cite{johnston-model}

The unique insulating property of BaMn$_{2}$As$_{2}$ makes it
promising to explore novel states of matter via applying high
pressures as well as various chemical doping just like
those\cite{122Fe,sefat,llj,js,hp} employed in BaFe$_{2}$As$_{2}$.
Some doping studies\cite{mis-gap,BaK} were recently performed in
BaMn$_{2}$As$_{2}$ system. For the Mn-site doping,\cite{mis-gap}
unfortunately, many transition metals do not substitute for Mn at
levels above 0.5\%. Although Cr and Fe (Mn's neighbors in the
Periodic Table) can substitute for Mn significantly (at levels of
4.4\% and 10\%, respectively), these substitutions hardly change the
electronic transport and magnetic properties. Very recently Pandey
et al.\cite{BaK} succeeded in doping small amount of K (1.6\% in
single crystals and 5\% in polycrystals) into BaMn$_{2}$As$_{2}$.
They demonstrated that the K doping induced hole carriers, and
changed the system into a metallic state. Interestingly the local
moment of Mn as well as the N\'{e}el temperature is basically
preserved against the K doping. It was thus advocated that the new
class of materials bridged the gap between the iron pnictide and
cuprate high $T_c$ materials.\cite{BaK} Another recent
study\cite{satya} showed a pressure-induced metallization in
BaMn$_{2}$As$_{2}$.

Motivated by the hole-doping result in LaMnAsO,\cite{syl} we also
succeeded in growing K-doped BaMn$_{2}$As$_{2}$ single
crystals.\cite{note} In this paper, we report the measurements of
electrical resistivity, magnetoresistivity, anisotropic
magnetization and heat capacity for the single crystals of heavily
doped Ba$_{1-x}$K$_{x}$Mn$_{2}$As$_{2}$ (BKMA) with $x$=0.19 and
0.26. The samples show metallic behavior with the resistivity one
order of magnitude smaller than previously reported\cite{BaK} in the
lightly K-doped samples. Besides, a weak ferromagnetic transition
was observed below $\sim$50 K on the background of local-moment AFM
order. Novel metallic state is also manifested by the
Curie-Weiss-like in-plane magnetic susceptibility, and the
appearance of a minimum in low temperature resistivity under high
pressures. Electronic structure calculations suggest that the
interplay between itinerant As 4$p$ holes and local Mn 3$d$ moments
plays an important role for the strange metallicity in BKMA system.

\section{\label{sec:level2}Experimental}
Single crystals of BKMA were grown by high-temperature spontaneous
nucleations using MnAs as the self flux. First, MnAs was prepared by
reacting stoichiometric mixture of Mn and As powder at 1073 K for 16
hours in an evacuated quartz tube. Second, Ba and K pieces with the
presynthesized MnAs powder were put into an alumina crucible in the
molar ratios of (Ba+K):MnAs=1:5, and K:Ba=0 and 1. All the starting
materials have the purity $\geq$99.9\%. For the growth of K-doped
crystals, the alumina crucible was placed into an iron tube sealed
by an iron cap, in order to protect the potassium from reacting with
quartz tube. All the procedures were conducted in a glove box filled
argon with the water and oxygen content below 0.1 ppm. The iron tube
was slowly heated up to 1503 K holding for 8 hours in a furnace
filled with protective argon atmosphere. Then it was allowed to cool
down to 1323 K at a cooling rate of 5 K/h, followed by shutting off
the furnace. Shiny black crystals with typical size of 2mm$\times$
2mm$\times$ 0.1mm were obtained. The as-grown crystals were found to
have two categories of K content. Those with higher K content
($x$=0.33$\sim$0.43) was inhomogeneous (e.g., they showed broad
x-ray diffraction and multiple weak ferromagnetic transitions), so
we will not discuss them in this paper. Others with lower K content
($x$=0.15$\sim$0.27) had homogeneous distribution of K. The two
crystals studied here were from the second category, and they were
carefully cleaved and selected to remove the flux MnAs as far as
possible. Note that the crystals had been stored in a glove box free
of oxygen and water at room temperature for 5 months.

X-ray diffraction was performed at room temperature using a D/Max-rA
diffractometer with Cu K$_{\alpha}$ radiation and a graphite
monochromator. As usual, the crystallographic $c$ axis was found to
be perpendicular to the crystal plate. Therefore, the lattice
parameter $c$ was able to be calculated by the (00$l$) multiple
diffractions after making the zero shift correction. The content of
the doped potassium in the single crystals were measured by
energy-dispersive x-ray spectroscopy (EDS), equipped in a
field-emission scanning electron microscope. The electrical
resistivity was measured by a standard four-probe method. Gold wires
were carefully attached to the crystal by using silver paint under
an optical microscope. The error caused by the size of the
electrodes leads to the total uncertainty in the absolute
resistivity of 10\%. Hydrostatic pressure was generated by a cubic
anvil cell immersed in a cryostat.\cite{cubic anvil} The magnetic
property was measured on a commercial Quantum Design magnetic
property measurement system (MPMS-5). The magnetization was measured
under $\mu_0$H=1 T in the zero-field cooling mode.

The electronic structure calculations were based on density
functional theory, using the CASTEP module of Materials
Studio.\cite{castep} The calculations employed ultrasoft
pseudopotentials and a plane-wave expansion for the wave functions.
The exchange correlation effects were treated by the generalized
gradient approximation with the Perdew-Burke-Ernzerhof functional.
The plane-wave basis energy cutoff was set at 330 eV. For the
calculation of 25\% K-doped BaMn$_{2}$As$_{2}$, a $2\times2\times1$
supercell was built. From the experimental result, $G$-type AFM
order was assumed to be the ground state. For bandstructure and
density-of-states (DOS) calculations, $26\times26\times15$ and
$5\times5\times3$ $k$-points were used, respectively.

\section{\label{sec:level3}Results and discussion}
\subsection{\label{sec:level1}Sample Characterizations}

\begin{figure}
\includegraphics[width=7.5cm]{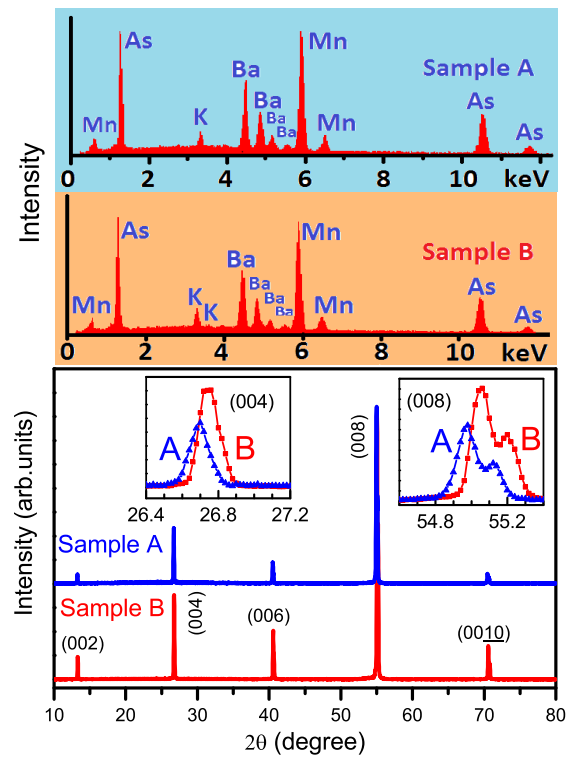}
\caption {(Color online) Compositional and structural
characterizations of the two Ba$_{1-x}$K$_{x}$Mn$_{2}$As$_{2}$
single crystals. The upper panels are the typical energy-dispersive
x-ray spectra which give $x$=0.19(1) and 0.26(1) for samples A and
B, respectively. The lower panel shows the x-ray diffraction
patterns of the identical samples. The the magnified (004) and (008)
reflections are displayed in the insets.}
\end{figure}

The BKMA single crystals were characterized by XRD and EDS
experiments. The typical EDS patterns for the samples studied here,
shown in the upper panels of Fig. 1, indicate the presence of only
four elements K, Ba, Mn and As. The quantitative analysis gives the
chemical formula as Ba$_{0.81(3)}$K$_{0.19(1)}$Mn$_2$As$_2$ and
Ba$_{0.74(3)}$K$_{0.26(1)}$Mn$_2$As$_2$ for Sample A (\$A) and
Sample B (\$B), respectively. The XRD patterns were recorded for the
plate-like crystals placed horizontally on a sample holder. The
diffraction peaks are very sharp [with the full width at half
maximum (FWHM) about 0.1$^\circ$], indicating good homogeneity and
crystallinity. They were indexed as (00$l$) reflections with
$l$=even, consistent with the body-centered tetragonal structure
(space group: $I4/mmm$). The crystallographic $c$ axes (13.350{\AA}
for \$A and 13.330{\AA} for \$B) are 0.77\% and 0.92\% smaller than
that of the undoped BaMn$_{2}$As$_{2}$ (13.454{\AA}). This
phenomenon is in contrast with the Ba$_{1-x}$K$_{x}$Fe$_{2}$As$_{2}$
system, which shows remarkable \emph{increase} in $c$ with the K
doping.\cite{122Fe} The possible reason is the strong $p-d$
hybridizations in BaMn$_{2}$As$_{2}$,\cite{band} resulting in the
shortening of Mn$-$As bondlength (and thus shrinkage of $c$ axes)
when doping holes in the $p-d$ valence bands.

\subsection{\label{sec:level2}Resistivity}

\begin{figure}
\includegraphics[width=7cm]{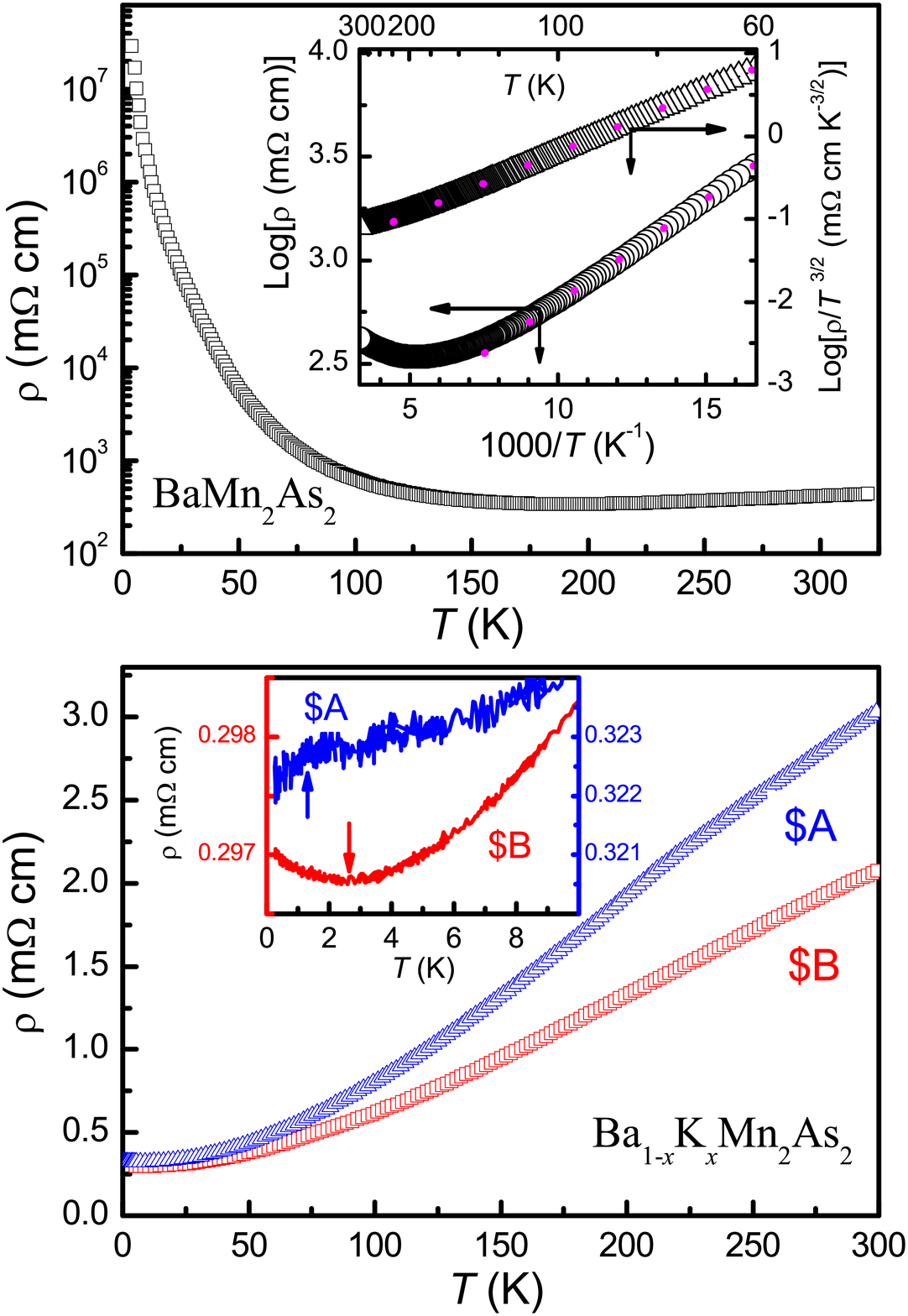}
\caption {(Color online) Temperature dependence of resistivity for
BaMn$_{2}$As$_{2}$ (upper panel) and
Ba$_{1-x}$K$_{x}$Mn$_{2}$As$_{2}$ (lower panel) crystals. The upper
inset plots log$\rho$ and log($\rho/T^{3/2}$) against the reciprocal
of temperature, corresponding to thermally-activated model and
non-adiabatic small-polaron transport model, respectively. The lower
inset shows the low-temperature resistivity. The K content of \$A
and \$B is 0.19(1) and 0.26(1), respectively.}
\end{figure}

Figure 2 shows the temperature dependence of electrical resistivity
[$\rho(T)$] of the undoped BaMn$_{2}$As$_{2}$ as well as the BKMA
crystals. The $\rho(T)$ data of the parent material indicate
semiconducting behavior with room-temperature resistivity of 430
m$\Omega \cdot$cm. However, the temperature coefficient is positive
(metallic-like) for $T> 190$ K. This result is qualitatively
consistent with those in previous reports.\cite{johnston-prop,band}
The absolute values of the resistivity are closer to those in
Ref.~\cite{band}. The data of 60 K$<T<$90 K can be well fitted with
the standard thermally-activated equation, i.e., $\rho \propto $
exp($E_{a}$/$k_{B}T$). The fitted result gives the activation energy
$E_a$=0.027 eV, precisely coinciding with the previous
report.\cite{johnston-prop} This suggests that the bandgap $E_g$ is
no less than 2$E_a$=0.054 eV, which agrees with the previous band
calculation result ($E_{g}\sim 0.1 $ eV).\cite{band} We note that
the $\rho(T)$ data in the high-temperature regime obey the
non-adiabatic small-polaron transport behavior, i.e., $\rho \propto
$ $T^{3/2}$exp($\Delta$/$k_{B}T$). The fitted $\Delta$ value is
0.040 eV (using the $\rho(T)$ data from 60 K to 160 K). Note that
$\Delta$=$E_{a}+E_{hop}$, therefore, the polaron-hopping potential
$E_{hop}$ is estimated to be 0.013 eV. This small-polaron transport
behavior suggests that the dominant activated charge carriers are
weakly localized instead of being itinerant in the parent compound.
It is noted that neither thermally-activated model nor small-polaron
one accounts for the low-temperature ($T<$ 60 K) data well. Instead
the electrical transport tends to be dominated by variable range
hopping (VRH) in the low-temperature regime, because the $\rho(T)$
data better fit the Mott's law of three-dimensional VRH,\cite{VRH}
i.e., $\rho \propto$ exp$(T_{0}/T)^{1/4}$.

In contrast to the semiconducting behavior above for undoped
BaMn$_{2}$As$_{2}$ crystals, the K-doped crystals show metallic
conduction in the whole temperature range, as shown in the lower
panel of Fig. 2. The room-temperature resistivity is 2$\sim$3
m$\Omega \cdot$cm, which is over two orders of magnitude smaller
than that of BaMn$_{2}$As$_{2}$, and about one order of magnitude
smaller than that reported for the lightly K-doped samples with
$x$=0.016 and 0.05.\cite{BaK} This further confirms
insulator-to-metal transition in BKMA. Below 1.5 K, \$A with
$x$=0.19 shows a tiny drop in resistivity. However, the resistivity
of \$B ($x$=0.26) exhibits an upturn below 2.8 K, like the
well-known Kondo effect.

We chose the $x$=0.19 sample to study its pressure effect. Under
high pressures, while the high-temperature resistivity decreases
rapidly, the low-temperature resistivity decreases subtly. By
careful examination, the low-temperature metallicity changes under
pressures. At $P$=2 GPa, the resistivity begins to \emph{increase}
with decreasing temperature below 10 K. This resistivity upturn is
enhanced at higher pressures. Apparently, the pressure effect
resembles that of the further K doping (see the above data of
$x$=0.26). The result contrasts with the pressure-induced
metallization in the parent compound BaMn$_{2}$As$_{2}$.\cite{satya}

\begin{figure}
\includegraphics[width=7.5cm]{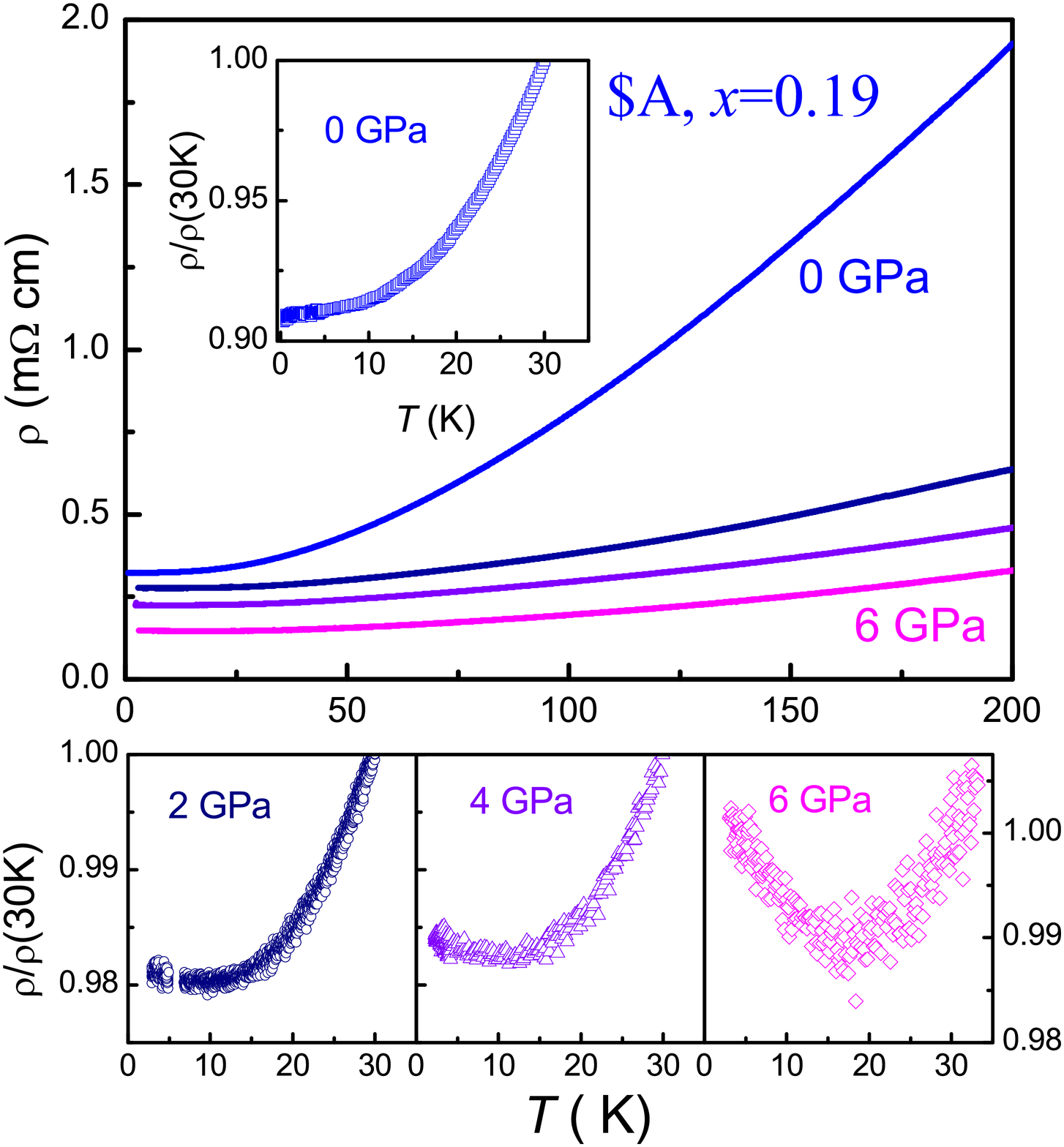}
\caption{(Color online) Temperature dependence of resistivity for
Ba$_{0.81}$K$_{0.19}$Mn$_2$As$_2$ under high pressures.}
\end{figure}

We note that the $\rho(T)$ data of $x$=0.05 sample can be fitted by
a power law $\rho=\rho_{0}+AT^{n}$ with $n$=2 in a wide temperature
range.\cite{BaK} However, the present heavily-doped samples have a
change in the power exponent below 50 K (see the upper panel of Fig.
4), which is associated with a weak ferromagnetic transition below.
The change in $n$ from $\sim$1.5 to $\sim$2.5 can be explicitly seen
in the series of plots with $T^{n}$ as the horizontal axes. The
transition temperature tends to increase with increasing K-content
or pressure.

\begin{figure}
\includegraphics[width=7.5cm]{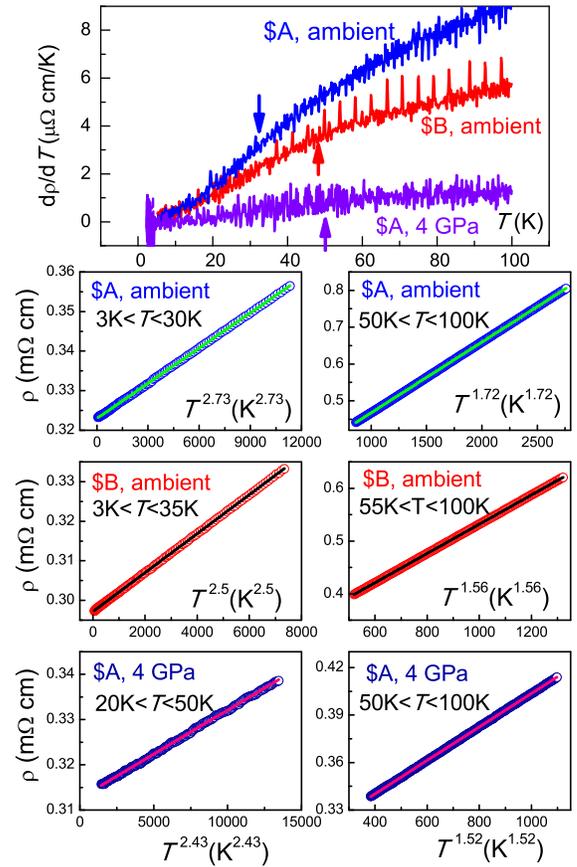}
\caption{(Color online) Upper panel: derivative of resistivity in
Ba$_{1-x}$K$_{x}$Mn$_{2}$As$_{2}$ ($x$=0.19 for\$A and$x$=0.26 for
\$B) single crystals. Other panels: plots of resistivity versus
$T^{n}$, indicating a change in the power exponent below $\sim$50
K.}
\end{figure}

\subsection{\label{sec:level3}Magnetization}

Figure 5 shows the magnetic susceptibility ($\chi$) of the
Ba$_{1-x}$K$_{x}$Fe$_{2}$As$_{2}$ single crystals. A remarkable
magnetic anisotropy can be seen from the difference in
$\chi_{ab}(T)$ and $\chi_{c}(T)$. The anisotropic ratio
$\chi_{ab}$/$\chi_{c}$ is over 10 below 100 K (see the inset of
figure 6), indicating the easy magnetization along the $ab$-plane.
The inset of Fig. 5 shows the $\chi_{c}(T)$ data separately. The
$\chi_{c}(T)$ data of both samples have a positive temperature
coefficient at high temperatures (the low-temperature upturn seems
to be related to the weak ferromagnetic transition to be discussed
below). This phenomenon is similar to that of the lightly K-doped
sample reported previously,\cite{BaK} which still has a $G$-type AFM
ground state with the Mn-moment along the $c$-axis. The
$\chi_{c}(T)$ behavior for $x$=0.19 and $x$=0.26 reflects that the
Mn local moments, oriented along the $c$-axis, are
antiferromagnetically ordered above room temperature. If so,
$\chi_{ab}(T)$ should be almost temperature-independent, as it is
for the parent compound,\cite{BaK} because the applied field is
perpendicular to the ordered moments.

\begin{figure}
\includegraphics[width=7cm]{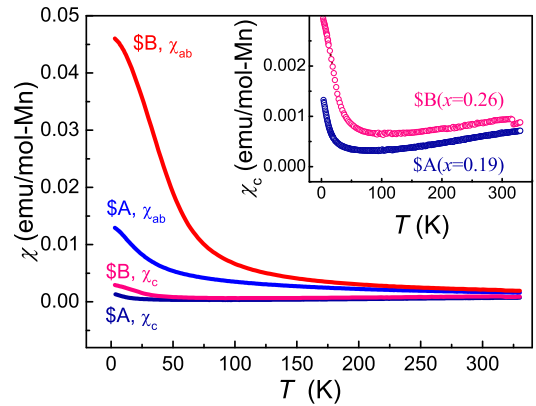}
\caption{(Color online) Temperature dependence of anisotropic
magnetic susceptibility for Ba$_{1-x}$K$_{x}$Mn$_{2}$As$_{2}$ single
crystals. The inset displays a magnified view for $\chi_{c}(T)$
($H\|c$). The small sudden increase in $\chi$ of the sample of
$x$=0.26 at 317 K is due to the presence of tiny amount
($\sim$0.01\%) of ferromagnetic MnAs impurity\cite{MnAs}.}
\end{figure}

However, the high-temperature $\chi_{ab}(T)$ data basically obeys
Curie-Weiss law, as shown in Fig. 6. One has to ascribe this
anomalous $\chi_{ab}(T)$ to the hole doping. This reminds us of the
Curie-Weiss metallic state in Na$_x$CoO$_2$,\cite{cw metal} where a
giant thermopower was rendered.\cite{terasaki,note2} By fitting the
$\chi_{ab}(T)$ data with an extended Curie-Weiss formula,
$\chi(T)=\chi_{0}+C/(T-\theta)$, we obtained the effective moment of
1.81$\mu_{B}$/Mn (1.94$\mu_{B}$/Mn) for $x$=0.26 ($x$=0.19),
equivalent to a local spin-1/2 paramagnet. The $\theta$ value for
$x$=0.26 is 32 K and $-$47 K for $x$=0.19. This result suggests a
magnetic exchange crossover from antiferromagnetic to ferromagnetic,
via tuning the potassium doping level.

\begin{figure}
\includegraphics[width=7.5cm]{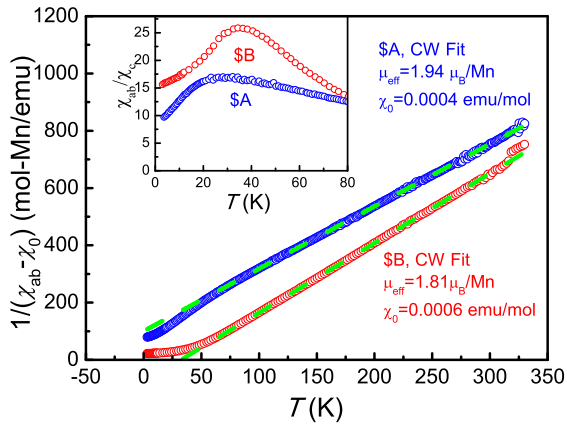}
\caption{(Color online) The reciprocal of the temperature-dependent
in-plane magnetic susceptibility, showing Curie-Weiss behavior in
Ba$_{1-x}$K$_{x}$Mn$_{2}$As$_{2}$ samples. The inset plots the
anisotropic ratio $\chi_{ab}(T)$/$\chi_{c}(T)$.}
\end{figure}

When $T<50$ K where the ratio $\chi_{ab}(T)$/$\chi_{c}(T)$ shows a
peak, $\chi_{ab}$ increases rapidly, or 1/$\chi_{ab}$ has a positive
curvature for both samples, especially for $x$=0.26. This suggests a
ferromagnetic component, which is confirmed by the $M_{ab}-H$ plot
showing in Fig. 7. At the temperatures above 55 K, the magnetization
increases linearly with the applied field. At low temperatures,
however, the $M_{ab}-H$ data show a jump in magnetization. The
saturated magnetization $M_{sat}$ is 0.02 $\mu_B$/Mn and 0.077
$\mu_B$/Mn for $x$=0.19 and $x$=0.26, respectively. A magnetic
hysteresis in the $M_{ab}-H$ loop is clearly seen for $x$=0.26,
further confirming the intrinsic weak ferromagnetism. Above 0.05 T,
the isothermal magnetization increases almost linearly and very much
slowly. Considered the large Mn local moment which is
antiferromagnetically ordered above room temperature,\cite{BaK} this
$M_{ab}(H)$ behavior suggests dominant local-moment
antiferromagnetism. The $M_{c}-H$ curve at 3 K for the $x$=0.26
crystal shows tiny saturated magnetization of 0.005 $\mu_B$/Mn (not
shown here). So, the spontaneous ferromagnetic moment is basically
along the basal planes. The absence of magnetic hysteresis and
smaller saturated magnetization for the sample of $x$=0.19 suggest
that it is on the verge of the emergent ferromagnetism.

\begin{figure}
\includegraphics[width=7.5cm]{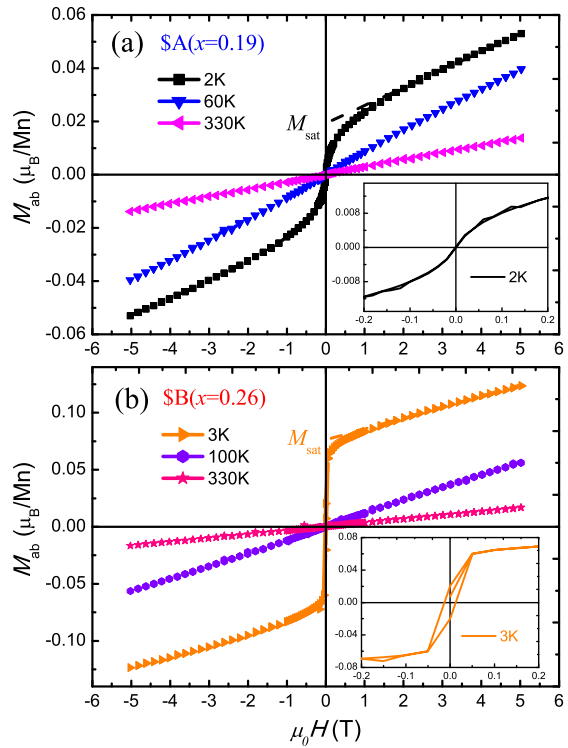}
\caption{(Color online) Isothermal $M_{ab}-H$ data at different
temperatures for samples A ($x$=0.19) and B ($x$=0.26). The Insets
show the enlarged $M_{ab}-H$ data at 2 K and 3 K.}
\end{figure}

\subsection{\label{sec:level4}Electronic Structure Calculation}

In order to understand the K-doping effect in BaMn$_{2}$As$_{2}$, we
employed first-principles electronic structure calculations. The
band structure of the undoped compound in Fig. 8(a) shows an
indirect band gap of 0.11 eV, which is consistent with the results
of above resistivity measurement and the previous
calculations.\cite{band,BaK} The projected DOS indicates strong
spin-dependent hybridization between Mn-3$d$ and As-4$p$ states,
which explains the reduced moment of 3.8 $\mu_B$/Mn. The As-4$p$ DOS
can be better accounted for the difference between the total DOS and
the Mn 3$d$ projection.\cite{122Cr} Thus we conclude that, compared
with the conduction bands, the valence bands have much more As-4$p$
weight. In this regard, the parent compound can be considered as a
charge-transfer insulator with strong Hund's rule coupling.

\begin{figure}
\includegraphics[width=7.5cm]{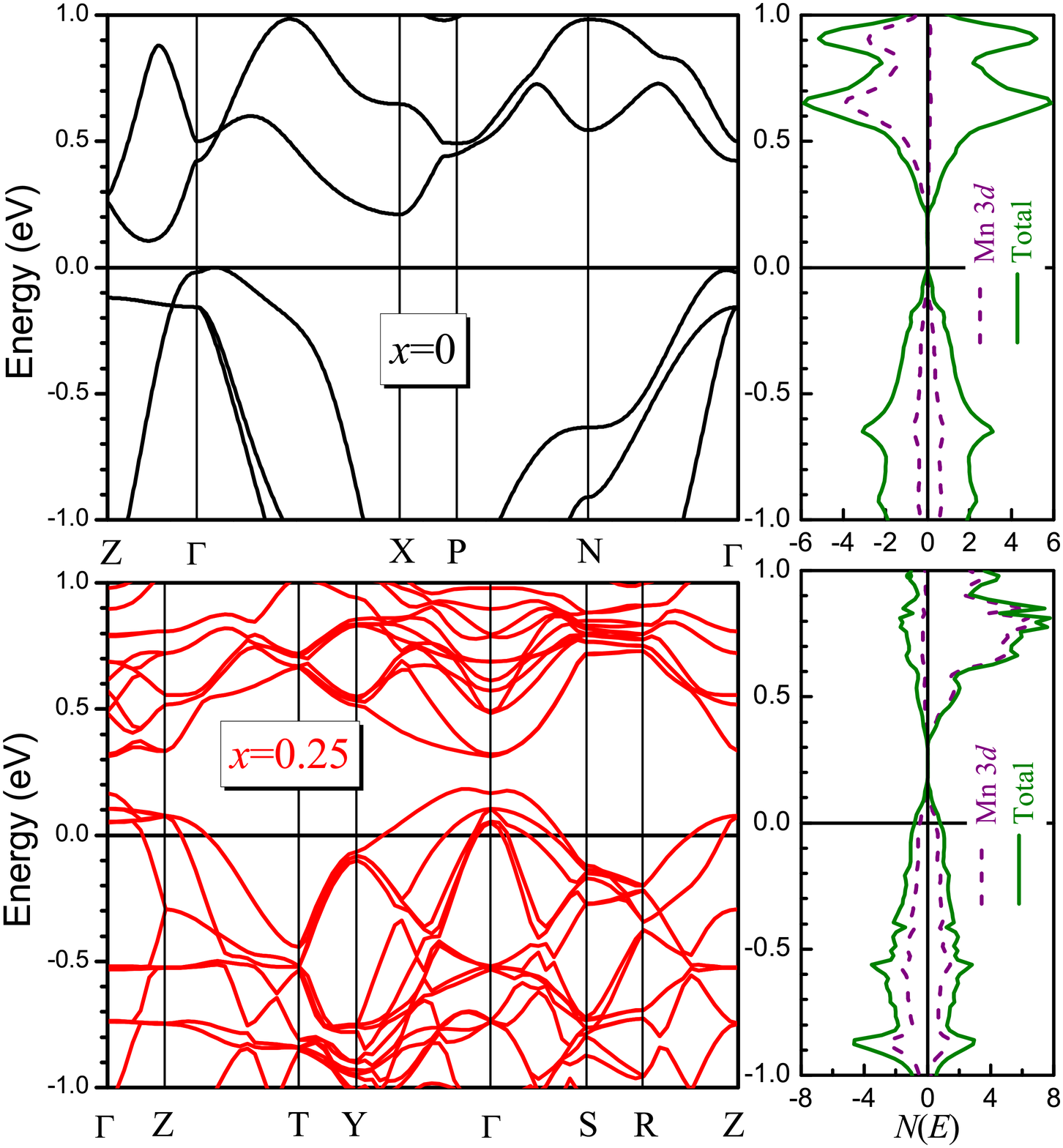}
\caption{(Color online) Band structures of BaMn$_{2}$As$_{2}$ (upper
panel) and Ba$_{0.75}$K$_{0.25}$Mn$_{2}$As$_{2}$ (lower panel). The
right panels show the corresponding total and projected (Mn 3$d$)
density of states (DOS). The negative DOS refer to the opposite
spin.}
\end{figure}

By using supercell approximation, we were able to calculate the
electronic structure of Ba$_{0.75}$K$_{0.25}$Mn$_{2}$As$_{2}$, which
is very close to the case of Sample B. As shown in Fig. 8(b), though
much more band dispersions appear near the Fermi energy due to the
use of supercell, the main features are maintained (e.g., a bandgap
of $\sim$ 0.1 eV still appears above $E_{F}$), except that the Fermi
level is shifted down by about 0.2 eV owing to the hole doping. The
shift of $E_{F}$ makes multiple hole-like Fermi surfaces, which
account for the metallic behavior in BKMA. Because As-4$p$ holes
usually have much higher mobility than the Mn-3$d$ holes do, the
metallic conductivity should be mainly contributed by the As-4$p$
holes. Therefore, the anomalous resistivity minimum under pressures
may be explained in terms of pressure-enhanced Kondo-like magnetic
scattering of the conduction holes.

Except for the rigid band scenario, the main modification in the
electronic structure of BKMA is that the As-4$p$ weight at $E_{F}$
is reduced, suggesting that the K doping weakens the $p-d$
hybridizations. Consistent with the reduced hybridization, the
calculated AFM local moment is significantly increased to 4.9
$\mu_B$/Mn, close to the value of a free Mn$^{2+}$ ion, and
consistent with the previous neutron diffraction result (4.2
$\mu_B$/Mn for $x$=0.05).\cite{BaK}

Now let us briefly discuss the possible physics of the anomalous
properties in BKMA. The weak ferromagnetism should be either local
or itinerant. In the local case, the ferromagnetism is easily
accounted for by the canting (toward the basal planes) of the Mn
local moment. The mechanism could be like the double exchanges in
manganites.\cite{DE} Note that the saturation magnetization is much
lower than the Mn local moment, therefore, such spin canting would
be very slight. The second is an itinerant ferromagnetism scenario,
in which the doped As 4$p$ holes and/or Mn 3$d$ holes are simply
spin-polarized. In this case, the saturation magnetization should be
comparable with the doping level per Mn ($x$/2), which basically
agrees with the experimental saturation moment. The Curie-Weiss-like
in-plane magnetic susceptibility resembles the behavior of
Na$_x$CoO$_2$ ($x\sim$0.7), as we pointed out above, suggests
correlated metallicity in the present system. As for the resistivity
minimum at high pressures, one has to consider the interactions
between conduction holes and the local moments, but we have not
arrived a definite picture so far. In a word, the interplay between
itinerant 4$p$ holes and local 3$d$ moments seems to be very
crucial. Future studies are called for to clarify all the novel
properties shown in Ba$_{1-x}$K$_{x}$Mn$_{2}$As$_{2}$ system.

\section{\label{sec:level4}Concluding remarks}

In summary, we have successfully synthesized heavily-doped
Ba$_{1-x}$K$_{x}$Mn$_{2}$As$_{2}$ ($x$=0.19 and 0.26) single
crystals which showed good homogeneity and crystallinity. By
measuring the resistivity and magnetization, we confirmed
metallization by a hole doping with potassium in BaMn$_{2}$As$_{2}$.
Several anomalous properties in the metallic state were
demonstrated. First, a weak ferromagnetic transition at $\sim$50 K
was discovered. Second, the low-temperature resistivity behaves a
Kondo-like minimum upon applying high pressures. Third, the in-plane
magnetic susceptibility shows Curie-Weiss behavior, though being in
an AFM state. Our electronic structure calculations indicate that
the As 4$p$ holes are most likely responsible for the metallic
conduction. Therefore, these novel metallic behaviors reflect the
interplay between itinerant 4$p$ holes and local 3$d$ moments in
Ba$_{1-x}$K$_x$Mn$_{2}$As$_{2}$ system.

\begin{acknowledgments}
This work is supported by the NSF of China (nos 11190023, 90922002
and 10934005) and the National Basic Research Program of China (nos
2010CB923003).
\end{acknowledgments}


\begin{thebibliography}{00}
\bibitem{note2}Unfortunately, our crystals are too small to make a reliable thermopower measurement. However, La$_{0.9}$Sr$_{0.1}$MnAsO indeed shows a very large thermopower at room temperature (Ref. 5).
\end{thebibliography}
\end{document}